\documentclass[pra]{revtex4}
\usepackage{graphicx}
\begin{document}
\title{Renormalization of the singular attractive $1/r^4$ potential}

\author{M. Alberg}
\affiliation{ Department of Physics, Seattle University, Seattle, WA
  98122, USA\\ and Department of Physics, University of Washington, Seattle, WA 98195, USA}
\author{M. Bawin}
\affiliation{Universit\'{e} de Li\`{e}ge, Institut de Physique B5, Sart
Tilman, 4000 Li\`{e}ge 1, Belgium  }
\author{F. Brau}
\affiliation{Groupe de Physique Nucl\'eaire Th\'eorique, Acad\'emie Universitaire Wallonie-Bruxelles, Universit\'e
de Mons-Hainaut, B-7000 Mons, Belgium}

\date{\today}

\begin{abstract}
We study the radial Schr\"odinger equation for a particle of mass $m$ in the field of a singular attractive $g^2/{r^4}$ potential with particular emphasis on the bound states problem. Using the regularization method of Beane \textit{et al.}, we solve analytically
the corresponding ``renormalization group flow" equation. We find in
agreement with previous studies that its solution exhibits a limit
cycle behavior and has infinitely many branches. We show that a
continuous choice for the solution corresponds to a given fixed number
of bound states and to low energy phase shifts that vary continuously
with energy. We study in detail the connection between this
regularization method and a conventional method modifying the short
range part of the potential with an infinitely repulsive hard core. We show that both methods yield bound states results in close agreement even though the regularization
method of Beane \textit{et al.} does not include explicitly any new scale in the problem. We further illustrate the use of the regularization method in the computation of electron bound states in the field of neutral polarizable molecules without dipole moment. We find the binding energy of $s$-wave polarization bound electrons in the field of C$_{60}$  molecules to be $17$ meV for a scattering length corresponding to a hard core radius of the size of the molecule radius ($\sim  3.37$ \AA). This result can  be further compared with recent two-parameter fits using the Lennard-Jones potential yielding binding energies ranging from $3$ to $25$ meV.  
 
\end{abstract}


\maketitle

\section{Introduction}
\label{sec1}

The renormalization of attractive singular potentials of the form $1/r^n$ with $n \geq 2$  was recently studied by Beane {\it et al.}~\cite{Beane}. We shall in the following refer to this renormalization method as the R-method. The purpose of this work is to analyze in more detail the case $n=4$ with particular emphasis on the bound states problem. On the physical side, this potential describes the long range part of the polarizability potential in atomic and molecular systems and is relevant to the description of the long range proton-deuteron electromagnetic interaction. From a more formal viewpoint, it is of interest to study how the regularization method of Ref.~\cite{Beane} for the $1/r^4$ potential compares with the results obtained for the $1/r^2$ potential \cite{Ba,Bra}, and whether it agrees with previous renormalization  schemes for the same interaction \cite{Muel,Kolo}. In this work, we follow Ref.~\cite{Ba} in order to find an analytic form of the solution to the renormalization group flow equation. We then compute both the bound states spectrum and the low energy phase shifts arising from the renormalized potential and compare the R-method to a conventional method using a hard core radius for regularizing purposes. We then use the R-method to discuss the binding energy of $s$-wave polarization bound electrons in the field of neutral molecules with zero dipole moment. Our main results are the following.
\begin{enumerate}
\item There are infinitely many solutions $\beta_n$, ($n =1, 2,
  3,\ldots$) to the renormalization group flow equation. Each
  $\beta_n$  exhibits a limit cycle behavior with, however, a period
  that depends on the cut-off radius $R$. Furthermore, $\beta_n$ takes the value $n\pi$ in the limit $R=0$.
\item A continuous choice of solution, obtained by jumping from one branch to the next closest branch below with decreasing values of the cut-off radius, corresponds to a renormalization with a given fixed number of bound states. However only the energy level with the weakest binding energy is insensitive to the value of the cut-off radius.
\item A numerical computation shows good agreements between the physical (in the sense discussed in Sec.~\ref{sec3}) bound states spectrum obtained with  the R-method and the corresponding spectrum obtained in a conventional method parameterizing the scattering length with a hard core radius. When applied to the problem of bound electrons in the field of polarizable molecules without dipole moment, we find in  particular that the R-method yields a binding energy of $17$ meV for $s$-wave polarization bound electrons in the field of C$_{60}$ molecules.
\end{enumerate} 

Our paper is organized as follows. In Sec.~\ref{sec2}, we present the
R-method proposed in Ref.~\cite{Beane} and we obtain the
renormalization group flow equation for the singular attractive
$1/r^4$ potential. In Sec.~\ref{sec3}, we discuss the bound states
spectrum of the regularized potential and discuss in  Sec.~\ref{sec4}
its connection with a conventional method modifying the short range
part of the potential by means of a hard core radius. In Sec.~\ref{sec5}, we discuss the application of  the R-method to the calculation of electron binding energies in the field of polarizable neutral molecules with zero dipole moment. In Sec.~\ref{sec6}, we show that the low energy phase shifts are, as expected, insensitive to the cut-off radius. Some concluding remarks are reported in Sec.~\ref{sec7}.

\section{Renormalization method}
\label{sec2}

In this paper, we follow the R-method proposed by Beane \textit{et al.} \cite{Beane} to obtain analytically the renormalization group behavior of the coupling constant of the short-range attractive square-well used to regularize the singular attractive $1/r^4$ potential.

We start with the $s$-wave reduced radial Schr\"odinger equation for one particle bound by a central potential $V(r)$ 
($\hbar= 2m=1$): 
\begin{equation}
\label{eq1}
\left( \frac{d^2}{dr^2} - V(r) - \kappa^2 \right) \psi(r) = 0,
\end{equation}
with $\kappa=\sqrt{-E}$ and where \cite{Beane}:
\begin{equation}
\label{eq2}
V(r) = -\frac{(\alpha_s)^2}{R^2} \theta (R-r) - \frac{(\alpha R)^2}{r^4} \theta(r-R)
\quad (\alpha_s, \alpha > 0),
\end{equation}
that is, the attractive $(\alpha R)^2/r^4$ is cut-off at a short distance  radius $R$ by an attractive square well. As in Ref.~\cite{Beane}, we first solve Eq.~(\ref{eq1}) for the zero energy solution $(\kappa=0)$ in order to find the corresponding wavefunction $\psi_0(r)$. This solution is given by:
\begin{eqnarray}
\psi_0 (r) &=& A \sin\left(\alpha_s\frac{r}{R}\right) \quad  r<R \\ 
\psi_0 (r) &=& B\ r\cos\left(\alpha\frac{R}{r} + \phi \right) \quad   r>R
\end{eqnarray}
where $\phi$ is the zero energy phase \cite{Beane} and is given by
\begin{equation}
\label{fi}
\tan \phi= L/g
\end{equation}
 where $L$ is the scattering length \cite{pere70} and $ g= \alpha R$.

\begin{figure}
\centering
\includegraphics*[width=8cm]{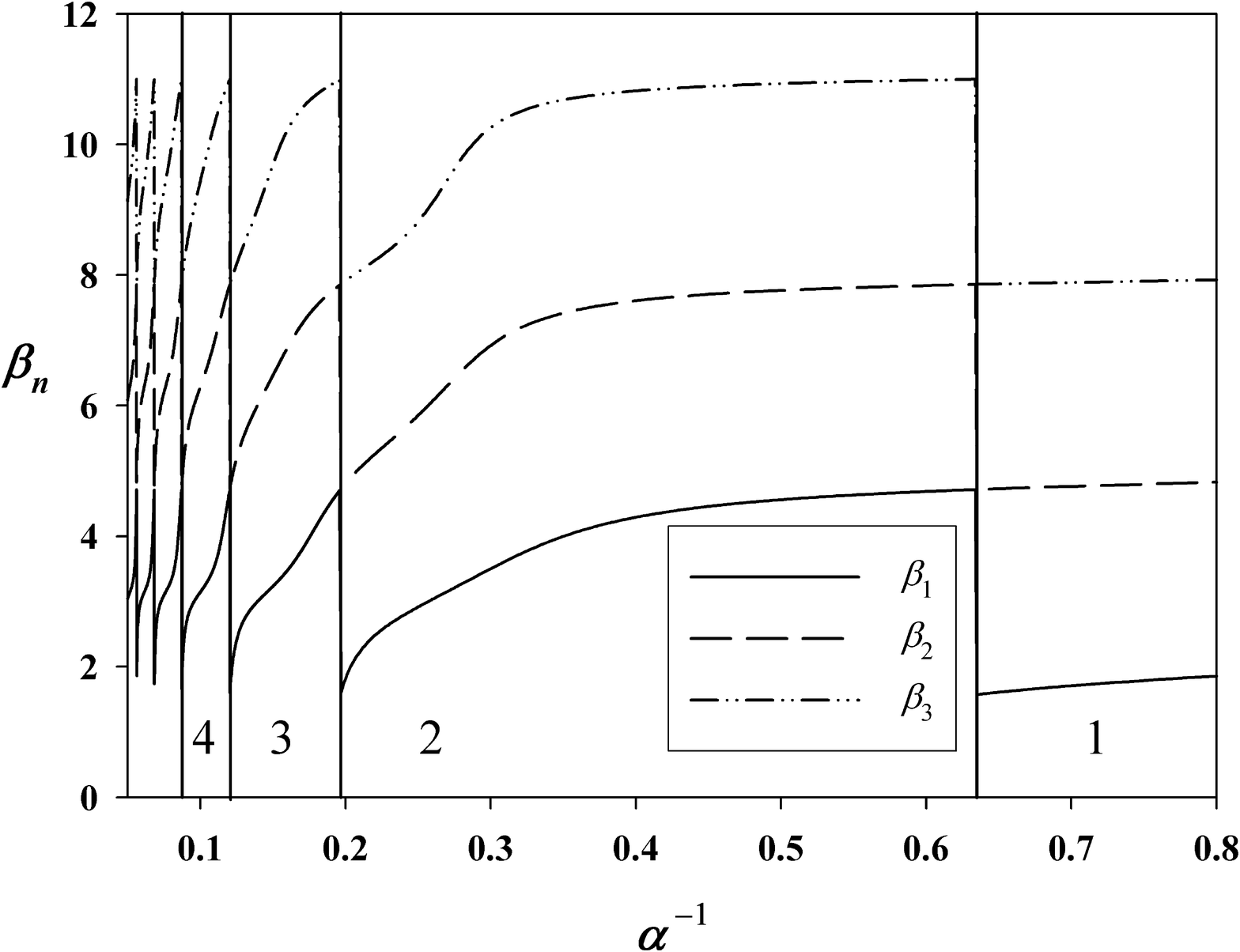}
\caption{The running coupling constant $\beta_n$ as a function of $ \alpha ^{-1}= R/g$ for $
n=1, 2, 3$ and $\phi = 1$. The regions labelled $i = 1,2,3,4$ are
discussed in Sec. III. Quantities on both axes are dimensionless.}
\label{fig1}
\end{figure}

The usual matching condition of the wave function and its derivative at $r=R$ then yields the renormalization group flow equation:
\begin{equation} \label{trans}
\alpha_s \cot \alpha_s = 1 + \alpha \tan (\alpha + \phi). 
\end{equation}
We can solve analytically Eq.~(\ref{trans}) as in Ref.~\cite{Ba} to obtain:
\begin{eqnarray}
\beta_0 &=& \pm \frac{{(\omega - 1)}^{1/2}}{\omega}\exp\left(\frac{1}{\pi}\int_0^1 \arg \Lambda_0(t)\frac {dt}{t} \right),
\quad \omega >1 \\
\beta_n &=& \pm n\pi \exp \left(\frac{1}{\pi}\int_0^1 \arg \Omega_n(t)\frac {dt}{t} \right),
\quad  -\infty<\omega < +\infty,\quad  n= 1,2,... 
\end{eqnarray}
where we denoted by $\beta _n $ the infinite set of solutions $\alpha _s $ and we have:
\begin{eqnarray}
\label{omega}
\frac{1}{\omega} &=& 1 + \alpha  \tan (\alpha  + \phi ), \\
\Lambda_0(t) &=& \lambda (t) + \textstyle \frac{1}{2}\omega t i \pi, \\
\lambda(t) &=& 1+\textstyle \frac{1}{2}\omega t \ln \frac {1-t}{1+t},\\
\Omega_n(t) &=& {\Lambda_0(t)}^2 + n^2{\pi}^2 {\omega}^2t^2. 
\end{eqnarray}
The integer $n$ is fixed on a given branch. The functions $\beta_n$
are given in Fig.~\ref{fig1} for $n=1,2,3$ as a function of $R$. For
computational ease, we chose $\phi = 1$. We only keep $n > 0 $
solutions as $\omega$ in formula (\ref{omega}) is unrestricted.
It appears that the coupling constant of the square-well potential is
a discontinuous function of $R$ for a given $n$. A similar behavior
was observed in the case of the renormalization of the singular
$1/r^2$ potential. Note however two important differences: on the one
hand, the period of oscillations, which changes with $R$, is no longer log-periodic \cite{Beane,Ba,Bra}; on the other hand, $\beta_n$ takes the value $n \pi$ in the limit $R\rightarrow 0$ for all $n$. Indeed, the coupling constant, $g=\alpha R$, of the $1/r^4$ potential is fixed for a given physical system (for example $g$ could be taken to be essentially the electrical polarizability, see Sec.~\ref{sec5}). Consequently, $\alpha$ varies with $R$ and is infinite in the limit $R\rightarrow 0$. This leads to a vanishing $\omega$, see (\ref{omega}), and $\Omega_n(t)$ is then real yielding $\beta_n = n\pi$ (the sign of $\beta_n$, and thus of $\alpha_s$ does not play a role, see (\ref{eq2})).

These results are consistent with the results of the study of the renormalization of long range attractive potentials in  Ref.~\cite{Muel}. As already discussed in Ref.~\cite{Beane}, one can also choose $\alpha_s$ to be a continuous function of $R$. This implies jumping from one branch of the solution to the next one just below at the point of discontinuity as illustrated in Fig.~\ref{fig1}. The respective merits of these two solutions were recently discussed in the literature in the case of the singular $1/r^2$ interaction~\cite{Ba,Bra}. 

\section{Bound states}
\label{sec3}

The renormalization method described in Sec.~\ref{sec2} only makes sense if the low energy observables (bound states and phase shifts for examples) are insensitive to the value of the cut-off radius $R$ (for $R$ small enough). The spectrum of the inverse square potential has been studied in detail \cite{Beane,Ba,Bra} and it was shown that bound states with an energy above $-1/R^2$ are indeed insensitive to the value of the cut-off.

As can be seen from Fig.~\ref{fig1}, discontinuities of $\beta_n$ appear for singularities of $\omega$ (see (\ref{omega})). For $R$ large enough, the function $|\alpha \tan(\alpha+\phi)|$ is smaller than 1 and no singularity can appear for $\omega$. The function $\beta_n$ is then continuous. This region is noted $1$ in Fig.~\ref{fig1}. With the formula giving the number of bound states of the regularized potential obtained in Appendix \ref{app1}, it is clear that in this region we have $n$ bound states if the coupling constant of the square-well is obtained with $\beta_n$ ($g=\phi=1$). In the region noted $2$ in Fig.~\ref{fig1}, formula (\ref{ap5}) shows that the number of bound states has increased by one upon crossing a discontinuity of $\beta_n$. Thus we must jump to a value $\beta_{n-1}$ of $\alpha_s$ in order to keep the number of bound states fixed. The various regions are obviously separated by the various discontinuities of $\beta_n$. In general, in region $i$, the potential has $n+i-1$ bound states for $\alpha_s=\beta_n$ ($g=\phi=1$). Obviously, the number of bound states in the region $i$ depends also on the value of $g$ and $\phi$ but this number always increases by one unit in the region $i+1$. 

\begin{figure}
\centering
\includegraphics*[width=8cm]{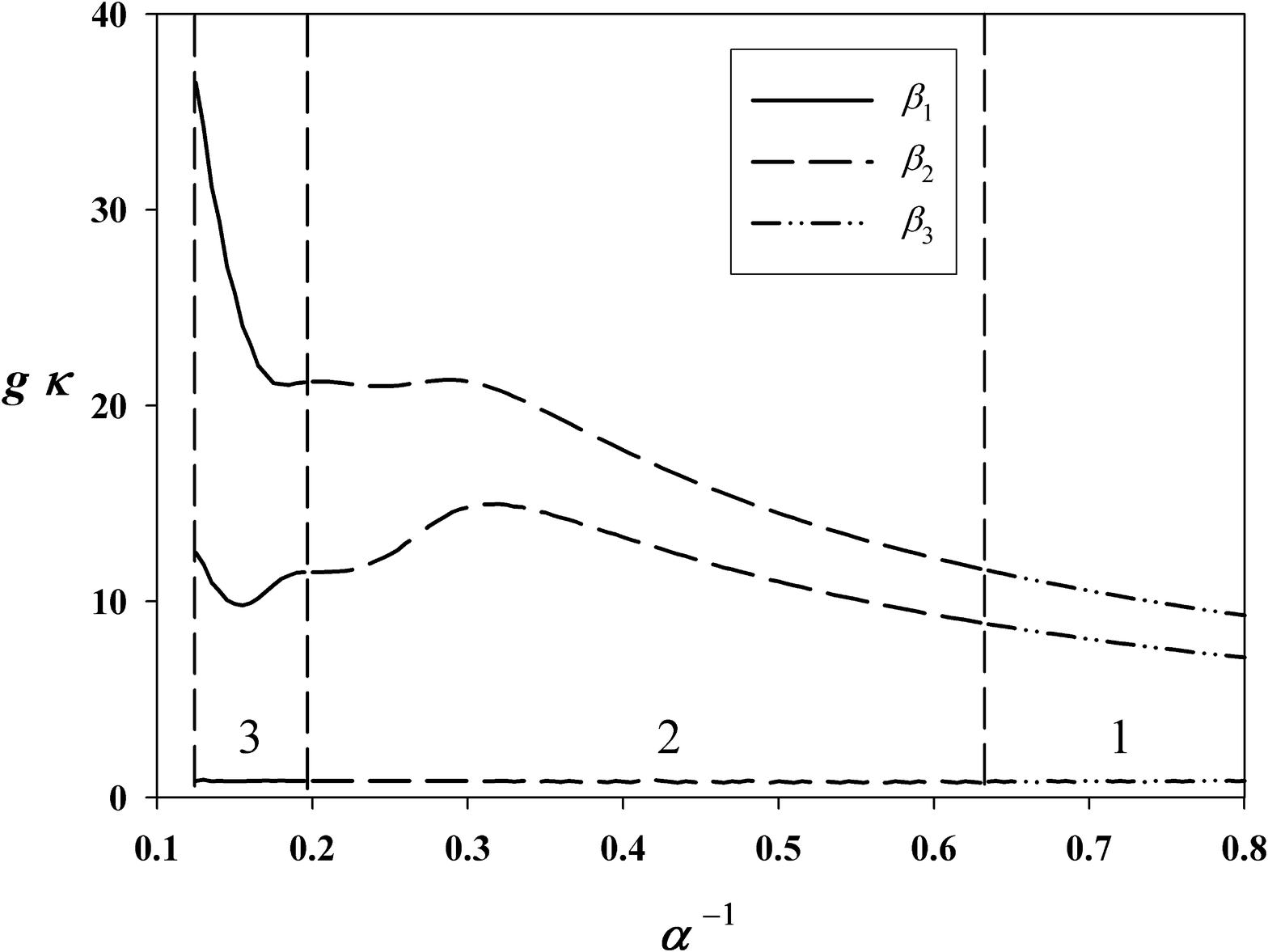}
\caption{The bound states spectrum as a function of $\alpha^{-1}= R/g$
  for a constant number of bound states fixed to three. Quantities on both axes are dimensionless.}
\label{fig2}
\end{figure}

With the analysis made above, it should be clear that a continuous choice of the solution $\alpha_s$ (hence mixing several branches with different values of $n$) corresponds to a renormalization with a fixed number of bound states. This is illustrated in Fig.~\ref{fig2}, where the evolution of the energy levels as a function of $R$ is given when three bound states are present in the regularized potential.

However, it is also clear that fixing the number of bound states in the potential yields a minimal value, $R_{\text{min}}$, for the cut-off radius $R$. For example, if we fix this number to $1$ ($g=\phi=1$), we must stay in the region $1$ of Fig.~\ref{fig1} with $\alpha_s$ computed with $\beta_1$. This leads to $R_{\text{min}}\approx 0.63\, g$.
A crucial outcome of our calculations as shown in Fig.~\ref{fig2} is the observation that the state with the weakest binding energy is insensitive to the value of $R$. The binding energy, as well as the mean square radius, of this state is also insensitive to the number of bound states present in the regularized potential. Since the states for which the binding energy varies with $R$ have no physical meaning and since the binding energy and the mean square radius of the state with the weakest binding energy are insensitive to $R$ and the number of bound states, there is no clear reason to choose a renormalization with a fixed number of bound states instead of a renormalization with a fixed branch since the latter does not introduce a minimal value for the cut-off radius $R$. Similar conclusions were drawn in the study of the singular attractive $1/r^2$ potential \cite{Ba,Bra}. We therefore conclude that the R-method used here yields bound states solutions independent of $R$ and independent of the particular branch of the solution to the renormalization group equation. We shall thus refer to these renormalized solutions as physical, and dismiss the deeply bound, $R$-dependent solutions as unphysical.

Now, according to our numerical analysis, the binding energies $E_{B}$ of these physical solutions for $\phi = 1$ are given by $E_{B} = \kappa^2$ with $\kappa$ given by  : 
\begin{equation}
\label{ph}
g \kappa \simeq 0.83.
\end{equation}
The form of formula (\ref{ph}) is counter-intuitive, as it
implies that the renormalized binding energy {\it increases} with decreasing $g$. This behavior is nevertheless simple to understand since a decrease of $g$ does not lead to an overall decrease of the regularized potential. The $1/r^4$ part of the regularized potential, see (\ref{eq2}), is indeed less attractive but the square-well part of the potential, which depends also on $g$, see (\ref{trans}), can be more attractive. The relation (\ref{ph}) is consistent with a WKB analysis of the Schr\"odinger equation as shown in Appendix \ref{app2}.

\section{Connection with hard core potentials}
\label{sec4}

It is of interest to compare results from the R-method to those
obtained with  conventional methods where the polarizability potential
$- \alpha^{P} e^2/(2r^4)$ is modified at the origin by means of some
short range repulsion \cite{Des,JMW} , $\alpha^P$ being the electric
polarizability of the system. An especially simple form of such a
modification  is the hard core regularization which implies that the bound particle wavefunction must vanish at some hard core radius ${\cal R}$. The attractive feature of this model is that the corresponding scattering length $L$ can be computed exactly to be \cite{ruskov}:
\begin{equation}
\label{rus}
L= \sqrt{\frac{\alpha^P}{a_0}}\cot\left(\sqrt{\frac{\alpha^P}{a_0}}{\cal R}^{-1}\right)
\end{equation}
where $a_0$ is the Bohr radius.
For a given value of $L$, Eq.~(\ref{rus}) gives the corresponding value of ${\cal R}$ for a given value of the electrical polarizability $\alpha^P$. On the other hand, the scattering length $L$ determines the value of $\phi$ from formula (\ref{fi}) in the R-method. Comparison between Eqs.~(\ref{fi}) and (\ref{rus}), and remembering that $\phi$ is defined modulo $\pi$, indicates that we have the correspondence
\begin{equation}
\label{cor}
{\cal R}=\frac{g}{(s+1/2)\pi-\phi},
\end{equation}
with $\phi \in [0, \pi]$ and $s=1,2,3,\ldots$. In Table \ref{tab1}, we show a comparison between the binding energies and root mean square radii obtained from both methods for the same values of the scattering length $L$, or equivalently for ${\cal R}$ and $\phi$ related by (\ref{cor}) with $s=1$.

\begin{table}
\protect\caption{Comparison between the binding energies $E_{B} ( \equiv \kappa^2)$ and the root mean square radii obtained with the R-method (with index R) and with the hard core regularization procedure (with index ${\cal R}$). The values of $\phi$ and ${\cal R}/g$ on the same line correspond to the same value of $L$ from formulas (\ref{fi}) and (\ref{rus}).}
\label{tab1}
\begin{center}
\begin{tabular}{cccccc}
\hline
$\phi$ & ${\cal R}/g$ & $(g \kappa)_{\text{R}}$ & $(g \kappa)_{{\cal R}}$ & $(\langle r^2 \rangle^{1/2}/g)_{\text{R}}$ & $(\langle r^2 \rangle^{1/2}/g)_{{\cal R}}$ \\
\hline
0.1 &  0.21681  &  3.09 &  3.14  &  0.545 & 0.548 \\
0.2 &  0.22161  &  2.73 &  2.82  &  0.577 & 0.583 \\
0.4 &  0.23189  &  2.18 &  2.23  &  0.666 & 0.672 \\
0.6 &  0.24317  &  1.69 &  1.71  &  0.794 & 0.796 \\
0.8 &  0.25560  &  1.23 &  1.24  &  0.982 & 0.984 \\
1   &  0.26937  &  0.830 & 0.834  & 1.300 & 1.300 \\
1.2 &  0.28471  &  0.484 & 0.486  & 1.960 & 1.961 \\
1.4 &  0.30190  &  0.196 & 0.196  & 4.175 & 4.176 \\
1.5 &  0.31130  &  0.0755& 0.0755 & 10.09 & 10.01 \\
\hline
\end{tabular}
\end{center}
\end{table}

We see from Table \ref{tab1} that the R-method and the hard core potential yield bound states values in excellent agreement. It is important to note, however, that this agreement holds for {\it physical} bound states, as defined above. This agreement also holds for any value of $s$. This integer is actually equal to the number of nodes of the wave function of the state with the weakest binding energy obtained with the hard core potential. However, there are significant differences between the two methods. The bound states wavefunction in the R-method has nodes corresponding to the number of unphysical (deeply bound) states and this number is actually arbitrary. The number of nodes of the hard core wavefunction is contrained by the value of ${\cal R}$ which should correspond to some characteristic length of the system. The correspondence between the two methods leads to a further understanding of relation (\ref{ph}). Indeed, ${\cal R}$ decreases with $g$ in (\ref{cor}) leading to an increase of the binding energy. Moreover, since the number of nodes of the wave function obtained with the R-method varies with $R$ (and the number of unphysical states), whereas the binding energy and mean square radius stay insensitive to this parameter, it is not easy to determine if the stable (physical) state is a ground state or an excited state. If the system considered has a given characteristic length, relations (\ref{rus}) and (\ref{cor}) can be used to determine $s$ (since $\phi$ is then known from (\ref{fi})) and then the position of the state in the spectrum. This is illustrated in the next section.

\section{Polarization bound states}
\label{sec5}

In this section we apply the renormalization R-method to the computation of weakly bound electron states in the field of polarizable neutral molecules with zero dipole moment. The C$_{60}$ molecule is one of the few possible candidates with such properties \cite{Des}. In the conventional approach, bound states are computed by solving the Schr\"odinger equation with a 2-parameter Lennard-Jones potential \cite{JMW}. Taking the electrical polarizability $\alpha^P$ of C$_{60}$ to be $558$ $a_0^3$ leads to an electron binding energy between $3$ and $25$ meV according to the value of the parameters describing the short range part of the interaction. We find a physical bound state at $17$ meV with $\phi = 1$ corresponding to a scattering length $L = g/\tan \phi$ with $g = \sqrt{m\alpha ^P e^2}=a_0 \sqrt{558}$. In a conventional hard core analysis, $\alpha^{P} e^2/2=279\, e^2 a_0^3$, the same binding energy is obtained with an effective radius of $6.37$ $a_0$ ($\sim  3.37$ \AA) for the C$_{60}$ molecule. This radius is obtained with the relation (\ref{cor}) with $s=1$. Larger values of $s$ yield the same binding energy but the effective radius has no longer a physical interpretation. Moreover, this weakly bound state is certainly a ground state \cite{JMW}. The value of the effective radius found with (\ref{cor}) is close to the experimental value of the mean radius of the C$_{60}$ molecule $3.55$ \AA\, \cite{krat90}. If this experimental value is used as effective radius, we find with (\ref{cor}) $\phi=1.192$ and the binding energy is then equal to $6$ meV (the relation (\ref{ph}) becomes for this value of $\phi$, $g \kappa \simeq 0.50$). Note that, as already clear from the results of Table \ref{tab1}, the binding energy depends strongly on the value of $\phi$. Thus the scattering length, which determines the value of $\phi$ (see (\ref{fi})) should be known with enough precision to allow definite predictions and comparisons with other regularization methods.

\section{Low energy phase shifts}
\label{sec6}

\begin{figure}
\centering
\includegraphics*[width=8cm]{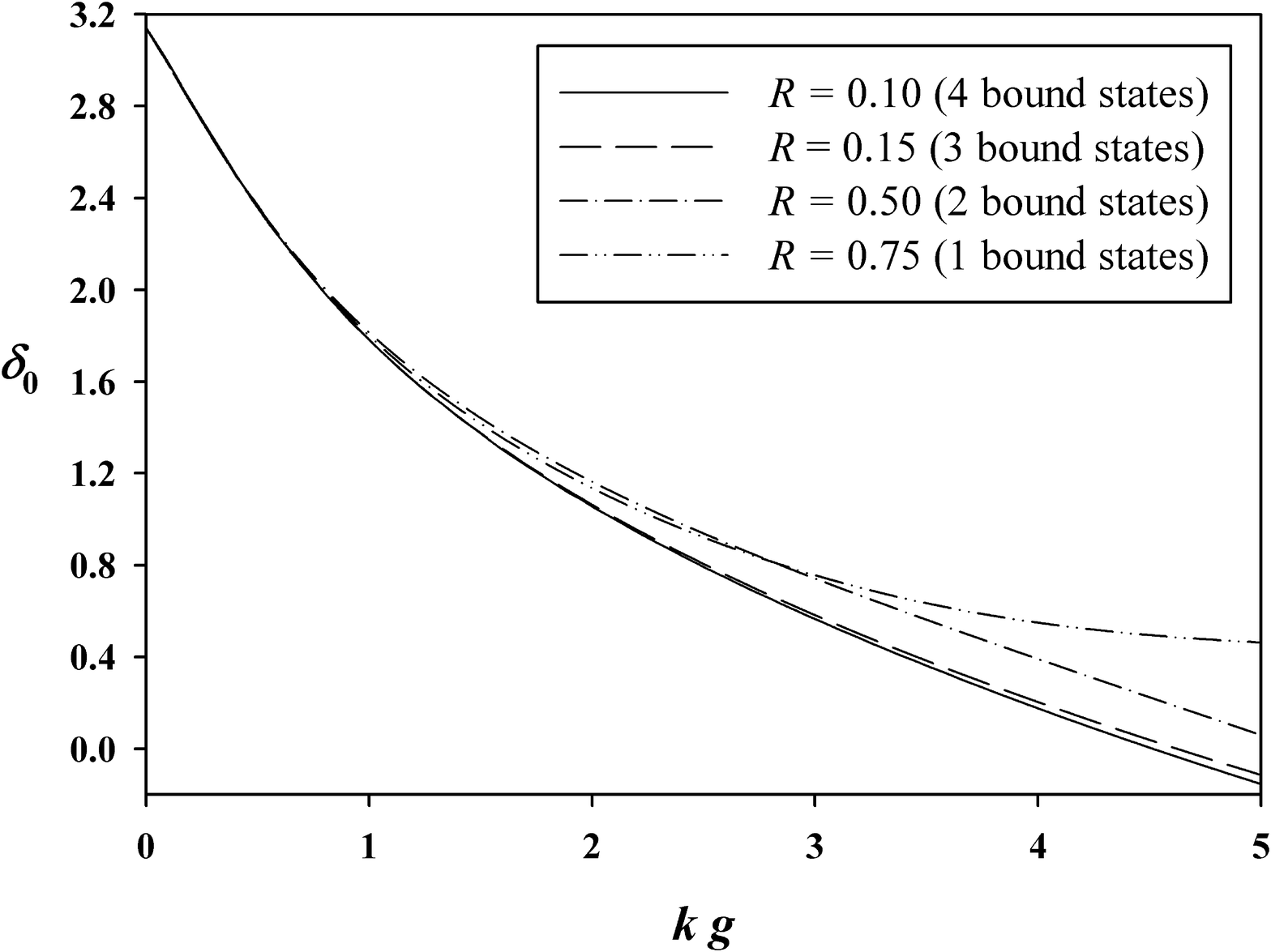}
\caption{The $s$-wave phase shift $\delta _0$ as a function of $gk$
  for various values of $R$ and for $\phi=1$. Quantities on both axes
  are dimensionless.}
\label{fig3}
\end{figure}

Another test of the method described in Sec.~\ref{sec2}, is the computation of the low energy phase shift. In Fig.~\ref{fig3}, we plot the $s$-wave phase shift $\delta_0$ as a function of $gk$ $(k \equiv \sqrt{E} > 0)$ for different values of $R$.
One can see again that $\delta_0$ is insensitive to the value of $R$ for $gk < 1$, even though the number of bound states (physical and nonphysical, as defined above) varies from $1$ to $4$. From the viewpoint of the renormalization method, only one (weakly bound) state is physical and leads to $\delta_0 = \pi$ for a vanishing energy.

\section{Conclusions}
\label{sec7}

In this paper we studied, with due emphasis on the bound states
problem, the renormalization of the singular $1/r^4$ potential using
the method of Beane {\it et al} \cite{Beane}. We found, in agreement
with previous works, that the solution $\beta_n$ to renormalization
group flow equation exhibits a limit cycle behavior and has infinitely many
branches. We discussed the form of the  bound states spectrum as a
function of $ \beta _n$ and discussed the connection between the
R-method and a conventional method using a hard core radius to modify
the short range part of the interaction. We then applied the R-method to calculate the energy of  polarization bound electrons to a neutral polarizable molecule without dipole moment. When applied to the C$_{60}$ molecule, we found that only very accurate values for the scattering length could discriminate between results from the R-method and those obtained from a polarization potential with hard core.

\appendix
\section{Number of bound states of the renormalized potential}
\label{app1}

It is well known that the number of bound states in a central potential is equal to the number of zeroes of the zero energy wave function in the interval $0<r<\infty$ (see for example \cite{calo67}). Equivalently, we count the number of extrema of this wave function. Thus we have to count the number of zeroes of the derivative of the zero energy wave function
\begin{eqnarray}
\label{ap1}
\psi_0'(r)&\sim& \cos \left(\alpha_s \frac{r}{R}\right)\quad \text{for}\quad r<R, \\
\label{ap2}
\psi_0'(r)&\sim& \cos \left(\frac{g}{r}+\phi\right)+\frac{g}{r}\sin\left(\frac{g}{r}+\phi\right)\quad  \text{for}\quad  r>R
\end{eqnarray}
The number of extrema, $N_1$, of the zero energy wave function in the interval $0<r<R$ is simply given
\begin{equation}
\label{ap3}
N_1=\left\{\left\{\frac{\alpha_s}{\pi}+\frac{1}{2}\right\}\right\},
\end{equation}
where $\left\{\left\{x\right\}\right\}$ is the integer part of $x$.
To find the number of extrema, $N_2$, of the zero energy wave function
in the interval $R<r<\infty$, we count the number of zeroes of the
expression $\cos(x+ \phi) + x \sin (x + \phi)$ for $x<\alpha$. Equivalently we search for the number of solutions of the equation
\begin{equation}
\label{ap4}
x \tan (x+\phi)=-1 \quad x<\alpha.
\end{equation}
The solutions are the intersection of a tangent function with a hyperbola. The number of solutions of the equation (\ref{ap4}) is given by 
\begin{equation}
\label{ap5}
N_2=\left\{\left\{\frac{1}{\pi}\left(\alpha+\phi+\arctan \frac{1}{\alpha}\right)\right\}\right\}.
\end{equation}
The total number is thus given by $N=N_1+N_2$ with $N_1$ and $N_2$ defined by (\ref{ap3}) and (\ref{ap5}) respectively.

\section{WKB analysis}
\label{app2}
The general formula giving the energy spectrum of a central potential in the WKB approximation is:
\begin{equation}
\int \limits_{r_-}^{r_+} dr \sqrt{E - V(r)} = (n -1/2) \pi,
\end{equation}
where $ n = 1, 2, 3,...$ and $r_{\pm}$ are the solution to $ E = V(r_{\pm})$. We then get from Eq.~(\ref{eq2}):
\begin{equation}
\label{w}
\int \limits_{0}^{R} dr \sqrt{E + \frac{(\alpha _s)^2}{R^2}} + \int \limits_{R}^{r_+} dr \sqrt{E + \frac{g^2}{r^2}} = (n -1/2) \pi.
\end{equation}
Performing the integrations in Eq.~(\ref{w}) and writing $ x = R/r_+$, we find:
\begin{equation}
\sqrt{ER^2 + (\alpha _s)^2} +\frac{g}{r_+}\left[-\frac{\sqrt{\pi}\, \Gamma (3/4)}{2\Gamma (5/4)} + x\sqrt{-1 + 1/x^4} + \frac{B(x^4,3/4,1/2)}{2} \right] =(n-1/2) \pi, 
\end{equation}
where $ B(x,a,b)$ is the Beta function. Assuming that $E$ remains
finite as $ R \rightarrow 0$ and using the formulas for $x \rightarrow 0$:
\begin{equation}
x\sqrt{-1 + 1/x^4} \simeq \frac{1}{x} -\frac{x^3}{2} \quad \text{and} \quad B(x^4,3/4,1/2) \simeq \frac{4x^3}{3},
\end{equation}
we eventually get, keeping the leading term as $ R \rightarrow 0$:
\begin{equation}
\label{wk}
g\sqrt{-E} = \frac{4}{\pi}\left[\frac{\Gamma (5/4)}{\Gamma (3/4)}
    \right]^2\left[\alpha _s + \frac{g}{R} - \pi(n - 1/2)\right]^2.
\end{equation} 
From Eq.~(\ref{ap3}) and Eq.~(\ref{ap5}), we find that $n\pi \simeq \alpha_s + g/R + \phi$ in the limit $ R\rightarrow 0$, so that we finally get from Eq.~(\ref{wk}):
\begin{equation}
\label{wkb}
g \kappa \equiv g\sqrt{-E} \simeq \frac{4}{\pi}\left[\frac{\Gamma (5/4)}{\Gamma (3/4)} \right]^2\left[\phi +  1/2\right]^2.
\end{equation}
Equation (\ref{wkb}) has the same functional form as Eq.~(\ref{ph}). It does not of course yield the same numerical value for the binding energy.

\begin{acknowledgments}
The work of M.B. and F.B. was supported by the National Fund for Scientific Research, Belgium. The work of M.A. was supported by the U.S. National Science Foundation under Grant No. 0245101.
\end{acknowledgments}

\end{document}